\documentstyle[aps,preprint]{revtex}
\tightenlines
\begin{document}
\title{Relative phase between the three-gluon and one-photon
amplitudes of the $J/\psi$ decays
\thanks{Parallel session talk at HADRON 2001, Protvino, Russia, August 30}}
\author{N.N. Achasov\\
Laboratory of Theoretical Physics, Sobolev Institute for
Mathematics\\
Academician Koptiug prospekt, 4, 630090 Novosibirsk,Russia}
\date{}
\maketitle

\begin{abstract}
It is shown that the study of the $\omega-\rho^0$ interference
pattern in the $J/\psi\to (\rho^0+\omega )\eta\to\pi^+\pi^-\eta$
decay provides  evidence for the large (nearly $90^\circ$)
relative phase between the isovector one-photon  and  three-gluon
decay amplitudes.
\end{abstract}
\vspace*{1in}

 In the last few years it  has been noted that the
single-photon and three-gluon amplitudes in the two-body
$J/\psi\to 1^-0^-$ and $J/\psi\to 0^-0^-$
\cite{castro-95,suzuki-98,suzuki-99}  decays appear to have
relative phases nearly $90^\circ$.

This unexpected result is very important to the observability of
CP violating decays  as well as to the nature of the $J/\psi\to
1^-0^-$ and $J/\psi\to 0^-0^-$ decays
\cite{castro-95,suzuki-98,suzuki-99,rosner-99,tuan-99,gerard-99,
tuan-01}. In particular, it points to a non-adequacy of their
description built upon the perturbative QCD, the hypothesis of the
factorization of short and long distances, and  specified wave
functions of final hadrons.  Some peculiarities of electromagnetic
form factors in the $J/\psi$ mass region were discussed in Ref.
\cite{achasov-98}.

The analysis \cite{castro-95,suzuki-98,suzuki-99} involved
theoretical assumptions relying on the strong interaction
$SU_f(3)$-symmetry, the strong interaction $SU_f(3)$-symmetry
breaking and the $SU_f(3)$ transformation properties of the
one-photon annihilation amplitudes. Besides,
 effects of the $\rho-\omega$ mixing in the
$J/\psi\to 1^-0^-$ decays were not taken into account in Ref.
\cite{castro-95} while
 in Ref. \cite{suzuki-98} the $\rho-\omega$ mixing was taken into
 account incorrectly , see the discussion in Ref. \cite{pisma}. Because of this, the model independent
determination of these phases are required.

Fortunately, it is possible to check the conclusion of Refs.
\cite{castro-95,suzuki-98} at least in one case
\cite{pisma,physrev}. We mean the relative phase between the
amplitudes of the one-photon $J/\psi\to\rho^0\eta$ and three-gluon
$J/\psi\to\omega\eta$ decays.

The point is that the $\rho^0-\omega$ mixing amplitude is
reasonably well studied
\cite{goldhaber-69,gourdin-69,renard-70,achasov-78,achasov-92,achasov-74,pdg-98}.
Its module and phase are known. The module of the ratio of the
amplitudes of the $\rho$ and $\omega$ production can be obtained
from the data on the branching ratios of the $J/\psi$-decays. So,
the investigation of the $\omega-\rho$ interference in the
$J/\psi\to (\rho^0+\omega )\eta\to\rho^0\eta\to\pi^+\pi^-\eta$
decay provides a way of measuring the relative phase of the
$\rho^0$ and $\omega$ production amplitudes.

Indeed, the $\omega-\rho$ interference pattern in the $J/\psi\to
(\rho^0+\omega )\eta\to\rho^0\eta\to\pi^+\pi^-\eta$ decay is
conditioned by the $\rho^0-\omega$ mixing and the ratio of the
amplitudes of the $\rho^0$ and $\omega$ production:
\begin{eqnarray}
\label{spectrum1} && \frac{dN}{dm}= N_\rho (m)\frac{2}{\pi}m\Gamma
(\rho\to\pi\pi\,,\, m)\left |\frac{1}{D_\rho (m)}\left
(1-\varepsilon (m)\left [\frac{N_\omega (m)} {N_\rho (m)}\right
]^{\frac{1}{2}}\exp\left\{i\left (\delta_\omega -
 \delta_\rho\right )\right\}\right ) \right. \nonumber\\[1pc]
&& \left. +\ \ \frac{1}{D_\omega (m)}\left (\varepsilon (m)+
g_{\omega\pi\pi}/g_{\rho\pi\pi}\right )\left [\frac{N_\omega
(m)}{N_\rho (m)}\right ]^{\frac{1}{2}} \exp\left\{i\left
(\delta_\omega - \delta_\rho\right )\right\}\right |^2
\end{eqnarray}
with
\begin{eqnarray}
\label{epsilon} && \varepsilon
(m)=-\frac{\Pi_{\omega\rho^0}(m)}{m_\omega^2-m_\rho^2 + im\left
(\Gamma_\rho (m)-\Gamma_\omega (m)\right )},
\end{eqnarray}
where m is the invariant mass of the $\pi^+\pi^-$-state, $N_\rho
(m)$ and $N_\omega (m)$ are the squares of the modules of the
$\rho$ and $\omega$ production amplitudes, $\delta_\rho$ and $
\delta_\omega$ are their phases, $\Pi_{\omega\rho^0}(m)$ is the
amplitude of the $\rho-\omega$ transition, $D_V(m)=m_V^2 - m^2 -
im\Gamma_V(m)$, $V=\rho$, $\omega$.

We obtained in Refs. \cite{pisma,physrev}
\begin{eqnarray}
\label{epsiloneff} && \varepsilon (m_\omega)+
g_{\omega\pi\pi}/g_{\rho\pi\pi}= ( 3.41\pm 0.24 )\cdot
10^{-2}\exp\left \{i\left ( 102\pm 1\right )^\circ\right \}\,.
\end{eqnarray}

The branching ratio of the $\omega\to\pi\pi$ decay
\begin{eqnarray}
\label{b} && B\left (\omega\to\pi\pi\right )=\frac{\Gamma \left
(\rho\to\pi\pi\,,\, m_\omega\right )}{\Gamma_\omega
(m_\omega)}\cdot\left |\varepsilon (m_\omega) +
g_{\omega\pi\pi}/g_{\rho\pi\pi}\right |^2.
\end{eqnarray}

The data \cite{markiii-88,dm2-90} were fitted with the function
\begin{eqnarray}
\label{fit} N(m) = L(m) + \left |\left (N_\rho\right
)^{\frac{1}{2}}F_\rho^{BW}(m) + \left (N_\omega\right
)^{\frac{1}{2}}F_\omega^{BW}(m)\exp\{i\phi\} \right |^2\,,
\end{eqnarray}
where $F_\rho^{BW}(m)$ and $F_\omega^{BW}(m)$ are the appropriate
Breit-Wigner terms \cite{markiii-88} and $L(m)$ is a polynomial
background term.

The results are
\begin{eqnarray}
\label{experiment} && \phi = ( 46 \pm 15)^\circ\,,\quad
N_\omega(m_\omega)/N_\rho =
8.86\pm 1.83\ \ \mbox{\cite{markiii-88}}\,,\nonumber\\
&& \phi = - 0.08\pm 0.17=(-4.58\pm 9.74)^\circ\,,\quad
N_\omega(m_\omega)/N_\rho =7.37\pm 1.72\ \ \mbox{\cite{dm2-90}}\,.
\end{eqnarray}

From Eqs. (\ref{spectrum1}), (\ref{b}), and (\ref{fit}) it follows
\begin{eqnarray}
\label{Nrho}
 && N_\rho = N_\rho (m_\rho)\left |1 - \varepsilon
(m_\rho)\left [ N_\omega(m_\rho)/N_\rho(m_\rho)\right
]^{\frac{1}{2}} \exp\{i\left(\delta_\omega - \delta_\rho\right
)\}\right |^2\,,\\
 \label{Nomega}
  && N_\omega =
B(\omega\to\pi\pi)N_\omega(m_\omega)\,,\\
\label{phi} && \phi =
\delta_\omega - \delta_\rho + arg\left [\varepsilon (m_\omega)+
g_{\omega\pi\pi}/g_{\rho\pi\pi} \right ] - \nonumber \\ && -
arg\left\{1 - \varepsilon (m_\rho)\left
[N_\omega(m_\rho)/N_\rho(m_\rho)\right ]^{\frac{1}{2}}
\exp\{i\left(\delta_\omega - \delta_\rho\right )\} \right\}\simeq
\nonumber\\ && \simeq \delta_\omega - \delta_\rho + arg\left
[\varepsilon (m_\omega)+ g_{\omega\pi\pi}/g_{\rho\pi\pi}\right ]
-\nonumber \\ &&- arg\left\{1 - \left |\varepsilon
(m_\omega)\right |\left [N_\omega(m_\omega)/N_\rho \right
]^{\frac{1}{2}}\exp\{i\phi\} \right\}\,.
\end{eqnarray}

From Eqs. (\ref{epsiloneff}), (\ref{experiment}) and (\ref{phi})
we get that
\begin{eqnarray}
\label{markiii} && \delta_\rho - \delta_\omega =
(60\pm 15 )^\circ\ \
\mbox{\cite{markiii-88}}\,,\\
\label{dm2} && \delta_\rho - \delta_\omega = 
(106\pm 10 )^\circ\ \ \mbox{\cite{dm2-90}}\,.
\end{eqnarray}

Whereas $\delta_\rho$ is the phase of the isovector one-photon
amplitude, $\delta_\omega$ is the phase of the sum of the
three-gluon amplitude and the isoscalar one-photon amplitude. But
luckily for us the latter is a small correction. Really, it
follows from the structure of the electromagnetic current
\begin{eqnarray}
\label{current}
 && j_\mu (x) =\frac{2}{3}\bar u(x)\gamma_\mu u(x) -
\frac{1}{3}\bar d(x)\gamma_\mu d(x) - \frac{1}{3}\bar
s(x)\gamma_\mu s(x) + ...
\end{eqnarray}
and the Okubo-Zweig-Iizuka rule the ratio for the amplitudes under
consideration ({\bf please image all possible diagrams!}):
\begin{eqnarray}
\label{sv}
 && \frac{A\left (J/\psi\to\mbox{ the isoscalar
photon}\to\omega\eta\right )}{A\left (J/\psi\to\mbox{ the
isovector photon}\to\rho\eta\right )\equiv A\left
(J/\psi\to\rho\eta\right )}=\frac{1}{3}\,.
\end{eqnarray}
Taking into account Eqs. (\ref{experiment}) and (\ref{Nrho}) one
gets
\begin{eqnarray}
\label{sg}
 && \frac{\left|A\left (J/\psi\to\mbox{ the isoscalar
photon}\to\omega\eta\right )\right|}{\left|A\left (J/\psi\to\mbox{
the three-gluon}\to\omega\eta\right )\right
|}\approx\frac{1}{9}\,.
\end{eqnarray}
From Eqs. (\ref{markiii}), (\ref{dm2}) and (\ref{sg}) one gets
easily for the relative phase ($\delta$) between the isovector
one-photon and  three gluon decay amplitudes
\begin{eqnarray}
\label{markiii1} && \delta = (60\pm 15 )^\circ\ - 4^\circ\ \ \ \ \
\mbox{\cite{markiii-88}}\,,\\
\label{dm21} && \delta = (106\pm 10 )^\circ - 6^\circ\ \ \ \ \
\mbox{\cite{dm2-90}}\,,
\end{eqnarray}
if the isoscalar and isoscalar  one-photon decay amplitudes have
the same phase. In case the isoscalar one-photon and  three-gluon
(isoscalar also!) decay amplitudes have the same phase
\begin{eqnarray}
\label{markiii2} && \delta = (60\pm 15 )^\circ\ \ \ \ \ \ \ \
\mbox{\cite{markiii-88}}\,,\\
\label{dm22} && \delta = (106\pm 10 )^\circ\ \ \ \ \ \ \
\mbox{\cite{dm2-90}}\,.
\end{eqnarray}

So, both the MARK III Collaboration \cite{markiii-88} and the DM2
Collaboration \cite{dm2-90}, see Eqs. (\ref{markiii1}),
(\ref{markiii2}) and (\ref{dm21}), (\ref{dm22}), provide support
for the large (nearly $90^\circ$) relative phase between the
isovector one-photon  and three-gluon decay amplitudes.

The DM2 Collaboration used statistics only half as high as the
MARK III Collaboration, but, in contrast to the MARK III
Collaboration, which fitted $N_\omega$ as a free parameter, the
DM2 Collaboration calculated it from the branching ratio of
$J/\psi\to\omega\eta$ using Eq. (\ref{Nomega}).

In summary I should emphasize that it is urgent to study this
fundamental problem once again with KEDR in Novosibirsk and with
BES in Beijing.

But I am afraid that only the $\tau$-CHARM factory could solve
this problem in the exhaustive way.

I gratefully acknowledge discussions with San Fu Tuan.

 The present work was
supported in part by the grant INTAS-RFBR IR-97-232.

\end{document}